\documentclass[12pt]{article}
\usepackage{epsfig}

\def\beq{\begin{equation}}
\def\eeq#1{\label{#1}\end{equation}}
\def\eeqn{\end{equation}}

\def\beqa{\begin{eqnarray}}
\def\eeqa#1{\label{#1}\end{eqnarray}}
\def\eeqan{\end{eqnarray}}

\let\bar=\overbar

\def\Dslash{\not{\hbox{\kern-4pt $D$}}}
\def\dslash{\not{\hbox{\kern-2pt $\del$}}}

\def\msb{{\bar{\ssstyle M \kern -1pt S}}}

\def\Title#1{\begin{center} {\Large {\bf #1} } \end{center}}

\begin{document}

\Title{Neutral and charged current cross section measurements and searches for new physics at HERA}

\bigskip\bigskip


\begin{raggedright}  
  Nicholas Malden\index{Malden, N.}\\
  High Energy Physics Group\\
  Schuster Laboratory\\
  The University of Manchester\\
  Manchester M13 9PL, U.K. \bigskip\bigskip
\end{raggedright}

\section{Introduction}

HERA is the only high energy electron--proton collider in the world
today and hence has unique opportunities both to probe the structure
of the proton and to search for physics beyond the Standard Model.
Results are presented for measurements of both neutral and charged
current cross sections, and for searches for exotic processes
involving direct electron--quark interactions (leptoquarks and
R--parity violating SUSY), generic coupling models (contact
interactions) and exclusive final states (isolated leptons and missing
$P_T$, single top production and pentaquarks).  Exclusion limits on
proposed models are set where no deviation from Standard Model
predictions are found.

\section{HERA, H1 and ZEUS}

At the HERA\cite{:1981uk} accelerator at DESY\footnote{Deutsches
  Elektronen Synchrotron} in Hamburg, Germany, positrons or electrons
have been collided with protons since the early 1990s. The
centre-of-mass energy of these collisions was $300\,\mathrm{GeV}$
prior to 1998, when a upgrade increased this energy to
$319\,\mathrm{GeV}$. These collisions take place at the heart of the
H1\cite{Abt:1993wz} and ZEUS\cite{:1993ee} experiments --
multi-purpose detectors with full solid angle coverage for tracking,
calorimetry and muon subdetectors.

\begin{figure}[htb]
\begin{center}
\epsfig{file=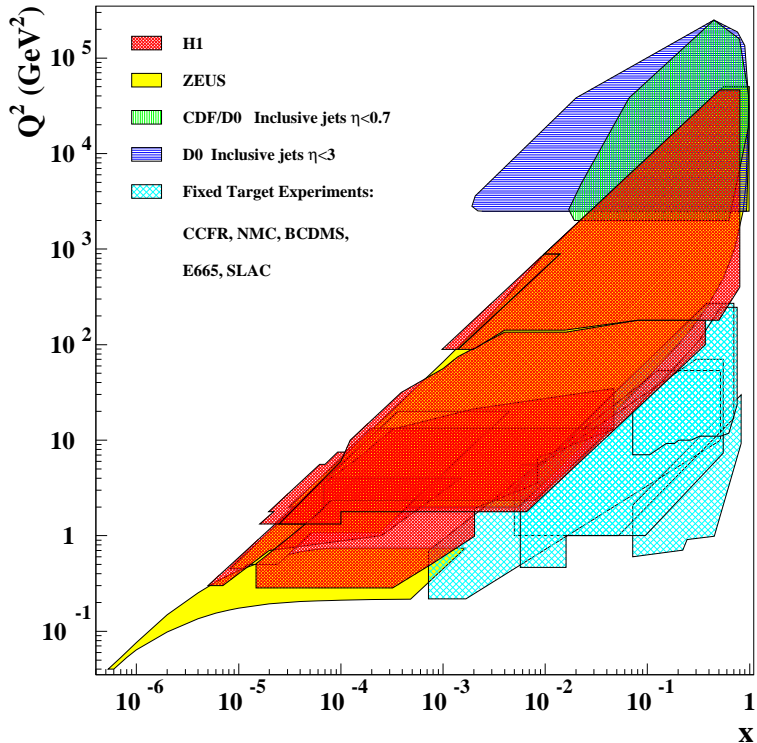,height=10cm}
\caption {The kinematic plane in $x$ and $Q^2$ accessible at HERA, compared to
  that of the Tevatron experiments and earlier fixed-target
  experiments.}
\label{fig:kinplane}
\end{center}
\end{figure}

By the end of the summer of 2000 an integrated luminosity of ${\cal
  O}(100)\,\mathrm{pb}^{-1}$ and ${\cal O}(15)\,\mathrm{pb}^{-1}$ had
been collected per experiment in $e^{+}p$ and $e^{-}p$ collisions,
respectively. The kinematic range covered by this data is illustrated
in figure \ref{fig:kinplane}, where $Q^2=-q^2$, $x=Q^2/(pq)$, and
$y=Q^2/sx$. Here $s$ is the centre-of mass energy squared, $q$ is the
four-vector of the exchanged boson, and $p$ is the four-vector of the
incoming proton.

\section{Neutral and Charged Current Cross Section Measurements}
\label{sec:nccc}

\begin{figure}[htb]
\begin{center}
\epsfig{file=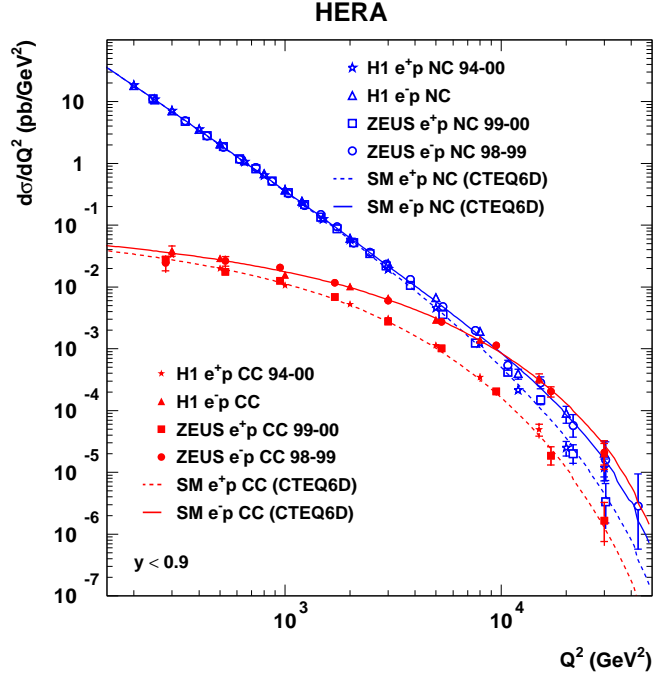,height=10cm}
\caption {The single-differential neutral current and charged current
  cross-section $\frac{d\sigma}{dQ^2}$, as measured by H1 and ZEUS in
  $e^{+}p$ and $e^{-}p$ collisions with $Q^2>200\,\mathrm{GeV}^2$.}
\label{fig:singlenccc}
\end{center}
\end{figure}

The single differential cross section $\frac{d\sigma}{dQ^2}$ for both
Neutral Current (NC) and Charged Current (CC) processes in $e^{+}p$
and $e^{-}p$ collision data is shown in figure \ref{fig:singlenccc}
for both H1\cite{h19900} and ZEUS\cite{zeusnc9900,zeuscc9900}.  The NC
cross-section is large at low $Q^2$ and decreases over several orders
of magnitude as $Q^2$ increases from $200\,\mathrm{GeV}^2$ to
$30000\,\mathrm{GeV}^2$. The $e^{+}p$ and $e^{-}p$ data overlay at low
$Q^2$, whereas at high $Q^2$ the $e^{+}p$ cross-section is smaller.
The CC cross-section is almost constant at low $Q^2$ but falls off
rapidly at high $Q^2$ and the $e^{+}p$ cross-section is significantly
lower than the $e^{-}p$ cross-section at high momentum transfer.  For
$Q^2>3000\,\mathrm{GeV}^2$ the NC and CC cross-sections are of similar
size. The data are well described by NLO QCD calculations.  The $Q^2$
dependence is governed by the propagator of the photon $\frac{1}{Q^4}$
for NC and by the propagator of the W-boson $\frac{1}{(Q^2+M_W^2)^2}$
for CC. The difference between $e^{+}p$ and $e^{-}p$ data is due to
the $Z^0\gamma$ interference in NC reactions. For CC interactions this
difference is attributed to the valence quark densities and the
relevant helicity factors.

When considering double differential cross sections it is convenient
to define the reduced cross section, which for the NC process is given by:\\

\large
\begin{center}
\begin{math}
 \tilde{\sigma}_{\mathrm{NC}}^{\pm}=
\frac{d^2\sigma_{\mathrm{NC}}}{dxdQ^2}/
\left(\frac{2\pi\alpha^2Y_{+}}{(Q^2)^2x}\right)\approx
F_2\mp\frac{Y_{-}}{Y_{+}}xF_3 
\end{math}
\end{center}
\normalsize

\begin{figure}[htb]
\begin{center}
\epsfig{file=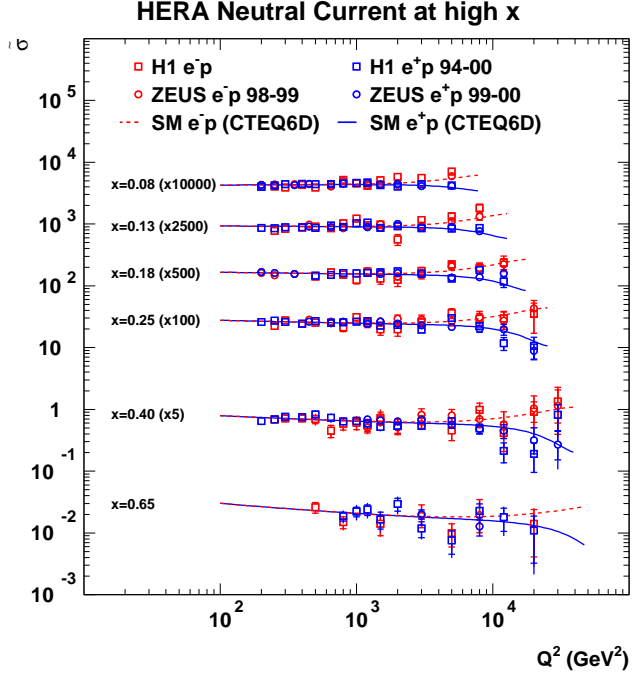,height=10cm}
\caption{The reduced neutral current cross-section $\tilde{\sigma}$,
measured by H1 and ZEUS in $e^{+}p$ and $e^{-}p$ collisions at high
$x$.}
\label{fig:doublenccc}
\end{center}
\end{figure}

~\\
where $Y_{\pm}=1\pm(1-y)^2$ and are known as the {\it helicity
  factors} and $F_2$ and $F_3$ are the proton structure functions.
The reduced cross-sections are shown in figure~\ref{fig:doublenccc},
measured for various values of $x$ as a function of $Q^2$.

The $Q^2$ dependence in this formalism is seen to be rather weak. The
main contribution to the cross-section is the photon exchange $F_2$.
The difference between $e^{+}p$ and $e^{-}p$ data at highest $Q^2$ is
due to $Z/\gamma$ interference. After subtracting the $e^{+}p$ data
from the $e^{-}p$ data, $xF_3$ may be measured at high $Q^2$. The
structure function $xF_3$ is sensitive to the valence quarks alone.

\begin{figure}[t]
\begin{center}
\epsfig{file=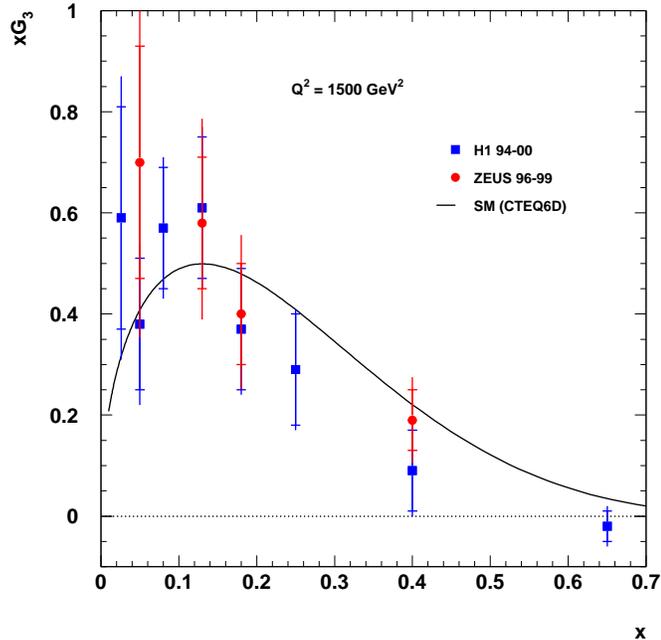,width=10cm}
\caption{The structure function $x\tilde{G}_3$, measured by H1 and ZEUS.}
\label{fig:xg3}
\end{center}
\end{figure}

Figure~\ref{fig:xg3} shows the HERA measurements of $x\tilde{G}_3$, where
propagator terms have been removed from $xF_3$. They compare well to
NLO QCD calculations.
~\\

In the same manner as for the NC cross sections it is convenient when
considering double differential CC cross sections to define the
reduced cross section $\tilde{\sigma}_{\mathrm{CC}^{\pm}}$:

\large
\begin{center}
\begin{math}
  \tilde{\sigma}_{\mathrm{CC}^{\pm}}=
\frac{d^2\sigma_{\mathrm{CC}}^{\pm}}{dxdQ^2}/
\left(\frac{G^2_FM_W^2}{(Q^2+M_W^2)^22\pi x}\right)
\end{math}
\end{center}
\normalsize

~\\
These reduced cross-sections are related to the quark densities,
$$\tilde{\sigma}_{\mathrm{CC}}^{-}\approx (xu++uc+x\bar{d}+x\bar{s}),$$
$$\tilde{\sigma}_{\mathrm{CC}}^{+}\approx (1-y)^2(xd+xs+x\bar{u}+x\bar{c})$$

\begin{figure}[t]
\begin{center}
  \epsfig{file=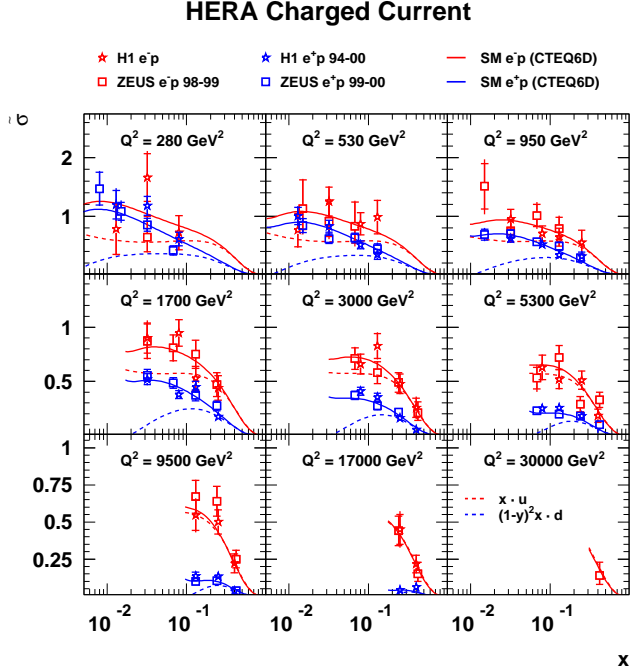,width=10cm}
\caption{The reduced charged current cross-section $\tilde{\sigma}$,
measured by H1 and ZEUS in $e^{+}p$ and $e^{-}p$ collisions at high
$x$.}
\label{fig:ccreduced}
\end{center}
\end{figure}

The HERA results are shown in figure~\ref{fig:ccreduced}. The data are
shown as a function of $x$ for nine values of $Q^2$.  The NLO QCD
calculation based on low energy data describes the data well.  Also
shown are the contributions from $u$ and $d$ valence quarks alone. The
CC data may be used to extract the $u$ ($d$) quark at high $x$ from
$e^{-}p$ ($e^{+}p$) data. For the $e^{+}p$ cross-sections the $d$
quark is suppressed by a helicity factor $(1-y)^2$.  A precise
determination of the valence quarks at highest $x$ requires a still
larger datasets, mostly in $e^{+}p$ collisions.

\section{Leptoquarks}
\label{sec:leptoquarks}

\begin{figure}[t]
\begin{center}
\epsfig{file=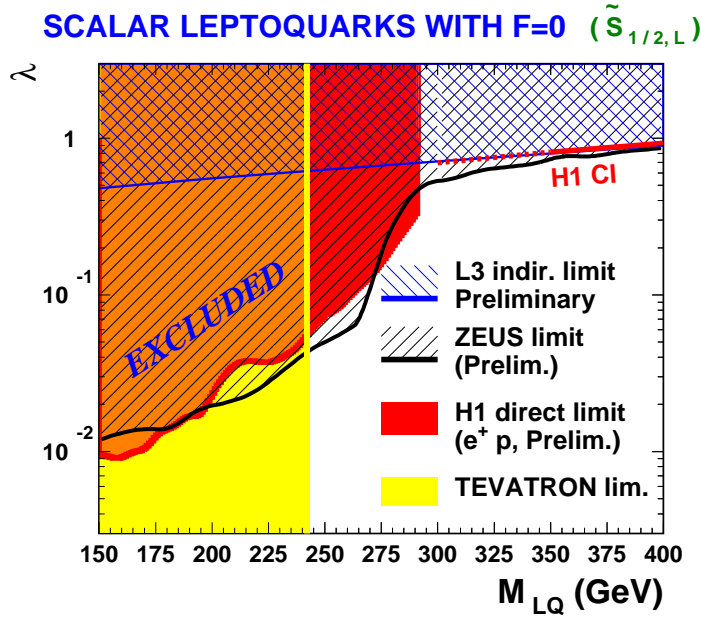,width=10cm}
\caption{Exclusion limits on the Yukawa coupling $\lambda$ as a function of leptoquark mass $M_{\rm LQ}$.} 
\label{fig:lq}
\end{center}
\end{figure}

Both Neutral Current (NC) and Charged Current (CC) high $Q^2$ data are
examined for evidence of leptoquark (LQ) production via either $s$ or
$u$ channel exchanges. This is done in the framework of the
BRW\cite{brw} model which predicts 7 scalar and 7 vector LQs. The $eq$
coupling is parameterised by the Yukawa coupling $\lambda$ and the
branching ratios are fixed. The data\cite{h1LQ,zeusLQ} show good
agreement with the Standard Model (SM) prediction and exclusion limits
in terms of $\lambda$ and LQ mass $M_{\rm LQ}$ are set. One such
result is shown in figure \ref{fig:lq}, with the complementary LEP and
Tevatron results.

A natural extension of this study is to look for evidence for lepton
flavour violation, mediated by $s$ or $t$ channel leptoquark exchange,
with subsequent decay into 2nd or 3rd generation leptons ({\it i.e.}
muons or taus). No evidence for such processes is observed and
exclusion limits are set in the $\lambda$-$M_{\rm LQ}$ plane within
the context of the BRW model. Taking a Yukawa coupling of
electromagnetic strength, couplings of scalar (vector) leptoquarks
with masses up to 275-300 (288-330) GeV to second generation leptons
and couplings of scalar (vector) leptoquarks with masses up to 260-284
(278-300) GeV to third generation leptons are excluded at 95\% C.L.
by H1\cite{h1LFV}.

A further extension of these models are contact interactions. These
models parameterise a coupling for the virtual exchange of particles
with masses beyond the direct access of the collider, but whose
interference with SM exchanges ($\gamma$, $Z^{\circ}$ and $W^{\pm}$)
could nevertheless be measureable. No deviations in the agreement of
the highest $Q^2$ NC data and the SM expectation are observed. These
results also set limits on finite quark radii. The ZEUS collaboration
set\cite{Chekanov:2003pw} an upper limit of $0.85 $ x $ 10^{-18}$m at 95\% C.L.

\section{Isolated Leptons and Missing {\boldmath $P_T$}}
\label{sec:isolep}
H1 has reported\cite{Andreev:2003pm,ichepisolep} an excess of events
containing an isolated electron or muon and missing transverse
momentum. Within the SM events of this topology are expected to be
mainly due to the production of a $W$ boson and its subsequent
leptonic decay, particularly when the hadronic system has high $P_T$
(large $P_T^X$).  Recent work\cite{nlo} has calculated the dominant
QCD corrections to the SM prediction at next-to-leading order (NLO).
The ZEUS Collaboration has also performed a search for such
events\cite{zeusisolep}. The results of these searches are presented
in table \ref{tab:isolep}.

\begin{table}[b]
\begin{center}
    \begin{tabular}{|c|c|c|c|}
      \hline
      H1 & Electrons & Muons & Combined \\
      94-00 $e^{+}p$ 163 pb$^{-1}$ & Obs'd/exp'd (sig.) & Obs'd/exp'd (sig.) & Obs'd/exp'd (sig.)\\
      \hline
      All data & 18 / 15.4$\pm$0.21 (71\%) & 9 / 4.1$\pm$0.7 (86\%) & 27 / 19.5$\pm$2.8 (74\%) \\
      \hline
      $P_T^X >$ 25 GeV & 8 / 2.6$\pm$0.5 (82\%) & 6 / 2.5$\pm$0.5 (88\%) & 14 / 5.1$\pm$1.0 (85\%) \\
      \hline
    \end{tabular}
    \begin{tabular}{|c|c|c|}
      \hline
      ZEUS preliminary & Electrons & Muons \\
      94-00 $e^{\pm}p$ 130 pb$^{-1}$ & Observed/exp'd (W) & Observed/exp'd (W) \\
      \hline
      $P_T^X >$ 25 GeV & 2 / 2.90$^{+0.59}_{-0.32}$ (45\%) & 5 / 2.75$^{+0.21}_{-0.21}$ (50\%) \\
      \hline
      $P_T^X >$ 40 GeV & 0 / 0.94$^{+0.11}_{-0.10}$ (61\%) & 0 / 0.95$^{+0.14}_{-0.10}$ (61\%) \\
      \hline
    \end{tabular}
\caption{Observed and expected number of events with an isolated electron or muon and missing transverse momentum for H1 (upper) and ZEUS (lower). The percentage of the SM expectation composed of signal processes ($W$ production) is also given for H1 (ZEUS).}
\label{tab:isolep}
\end{center}
\end{table}

\begin{figure}[ht]
  \epsfig{file=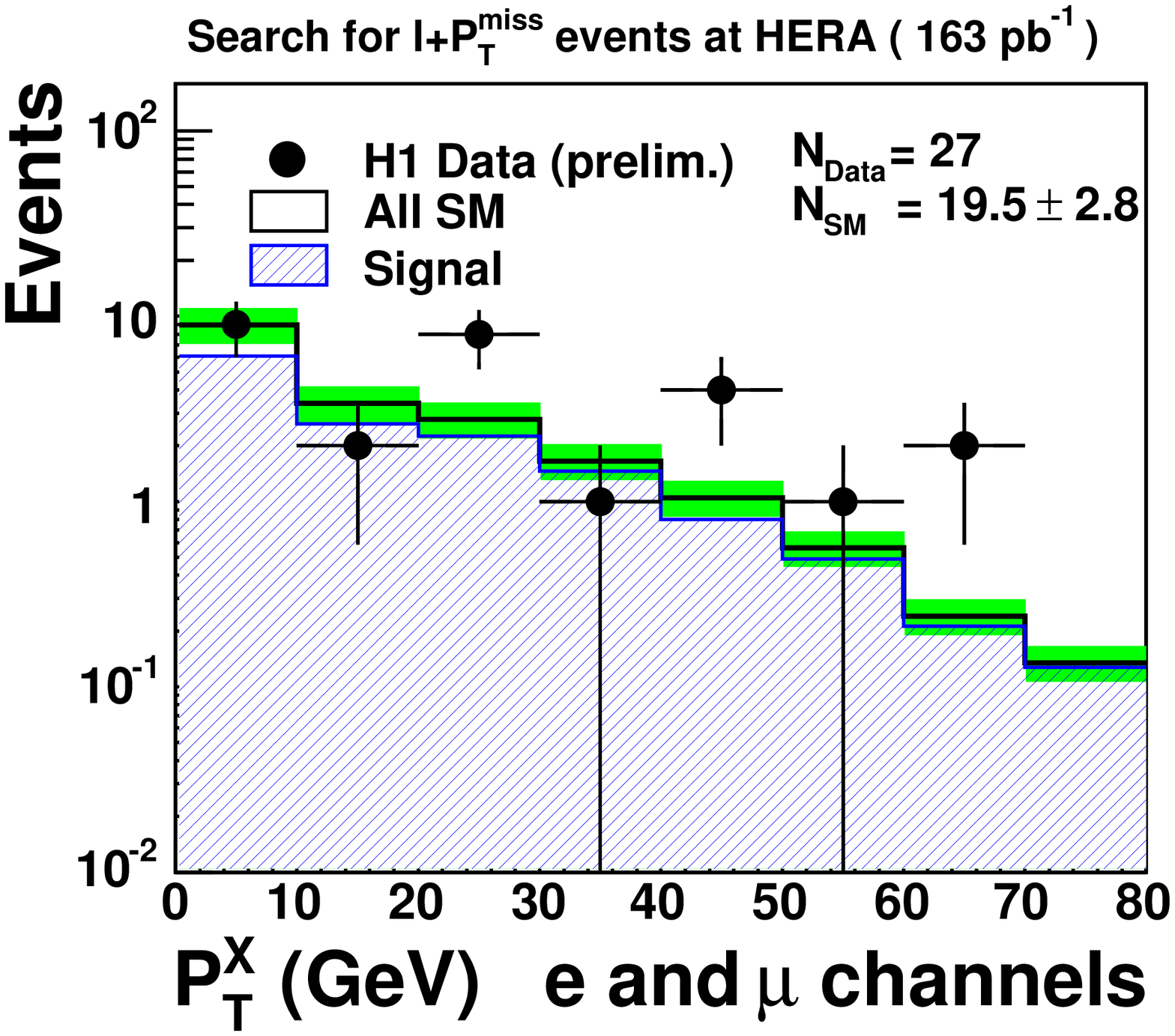,height=7cm}
  \epsfig{file=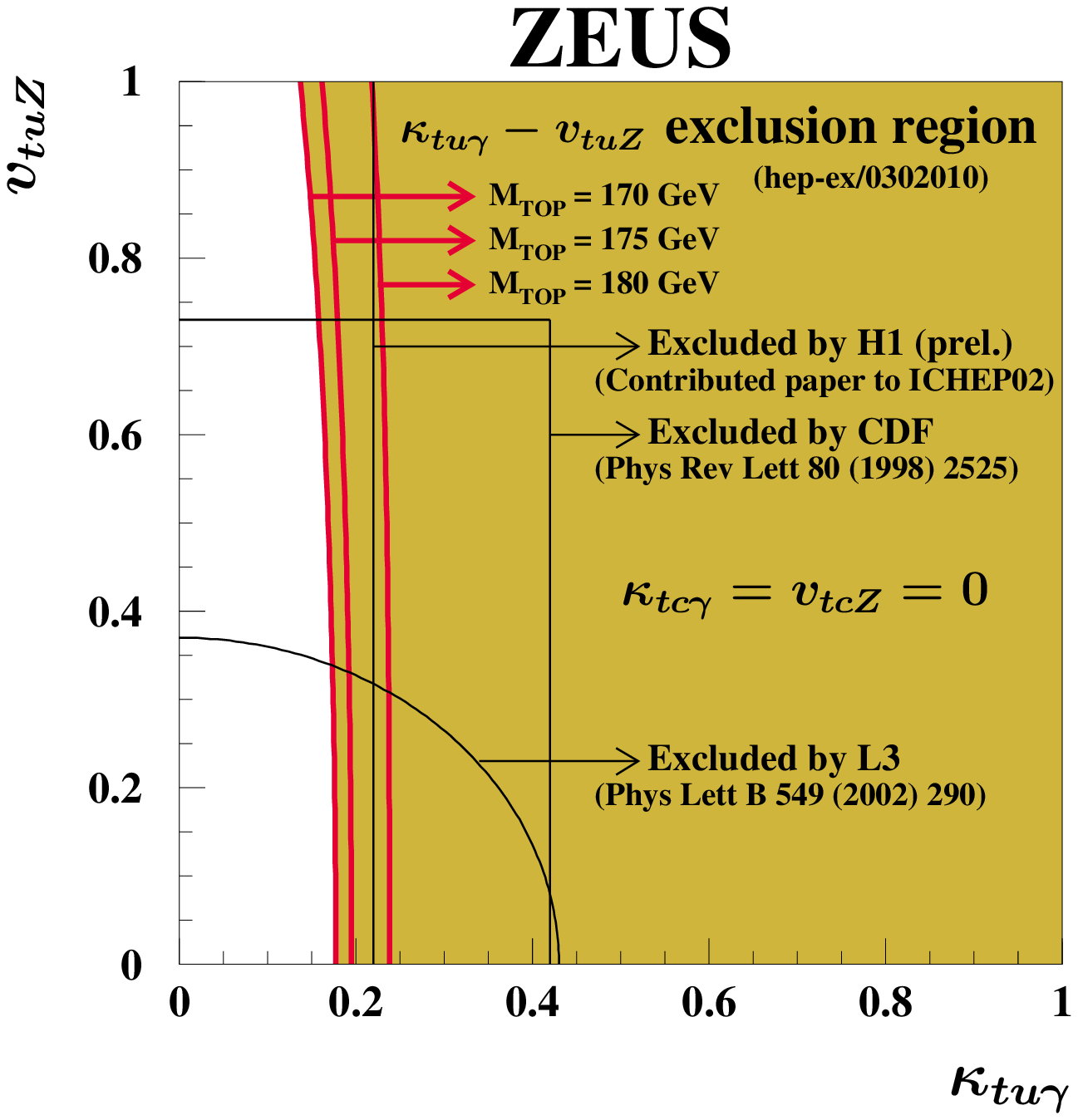,height=7cm}
  \caption{(Left) Number of events with isolated leptons (electrons or muons) and missing transverse momentum as a function of $P_T^X$, the transverse momentum of the hadronic system. (Right) Excluded regions of the anomalous coupling $\kappa_{t u \gamma}$-$v_{t u Z}$ plane.}
  \label{fig:isolep}
\end{figure}

The number of events with an isolated electron or muon observed by H1
overshoots the SM prediction, in particular at high $P_T^X$. The
distribution of events observed by H1 is shown in figure
\ref{fig:isolep} (left) with respect to $P_T^X$. Additionally, the
ZEUS Collaboration has searched in the tau channel\cite{zeusisotau},
finding 2 candidate events at $P_T^X >$ 25~GeV compared to a SM
expectation of $0.12\pm0.02$.
 
An event topology of an isolated lepton, missing $P_T$ and a high
$P_T$ hadronic jet may also be the signature of single top production,
where the top quark decays to a $b$ quark and a $W$. The rate of this
process is, however, negligible in the SM, due to the flavour changing
neutral current (FCNC) vertex required. The anomalous coupling at the
two relevant vertices $t u \gamma$ and $t u Z$, is parameterised by
the magnetic coupling $\kappa_{t u \gamma}$ and the vector coupling
$v_{t u Z}$ respectively. Both collaborations have also searched for
hadronic decays of single top quarks, but the large background from
other multi-jet processes severely restricts the contribution of this
channel to the analysis. The combined
results\cite{zeusisolep,h1singtop}, in terms of exclusion limits for
the anomalous couplings $\kappa_{t u \gamma}$ and $v_{t u Z}$, are
shown in figure \ref{fig:isolep} (right).

\section{R-Parity Violating SUSY}
\label{sec:rpsusy}

\begin{figure}[htb]
\begin{center}
\epsfig{file=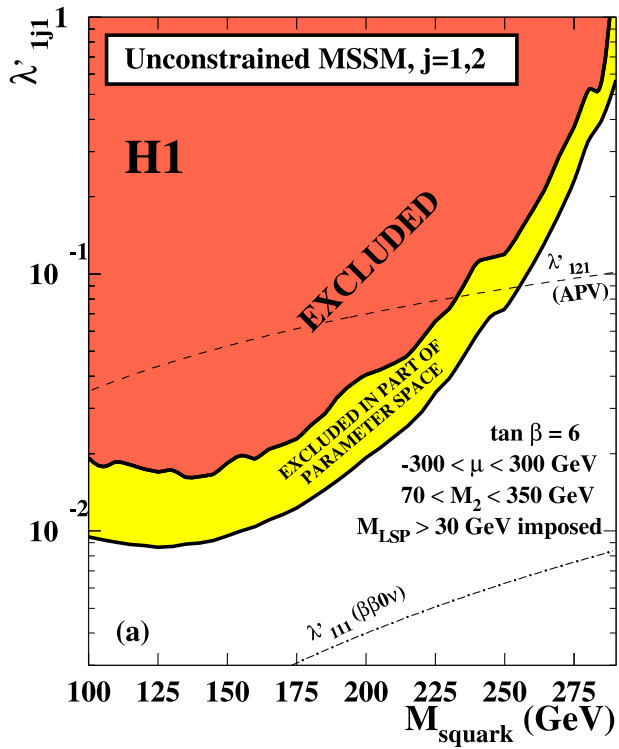,height=10cm}
\caption{Exclusion limits in R-parity violating SUSY searches on the Yukawa coupling as a function of squark mass.}
\label{fig:rpsusy}
\end{center}
\end{figure}

Since R--parity ($R_p$) is even (+1) for all SM particles and odd (-1)
for their supersymmetric (SUSY) partners, its violation implies that
SUSY particles may be singly produced and that the lightest SUSY
particle (LSP) is not stable. Resonant squark production at
HERA\cite{rpsusyherbi} is searched for in the framework of both the
minimal SUSY SM (MSSM) and the minimal supergravity (mSUGRA) models.
Some cascade decays result in background-free channels. No
evidence\cite{h1rpsusy,zeusrpsusy} for such processes is found
allowing mass and coupling limits to be set with the free variation of
the MSSM parameters $\mu$, $M_2$ and $\tan \beta$. One such result is
shown in figure \ref{fig:rpsusy}.

\section{Pentaquark searches}
\label{sec:pq}

\begin{figure}[ht]
  \epsfig{file=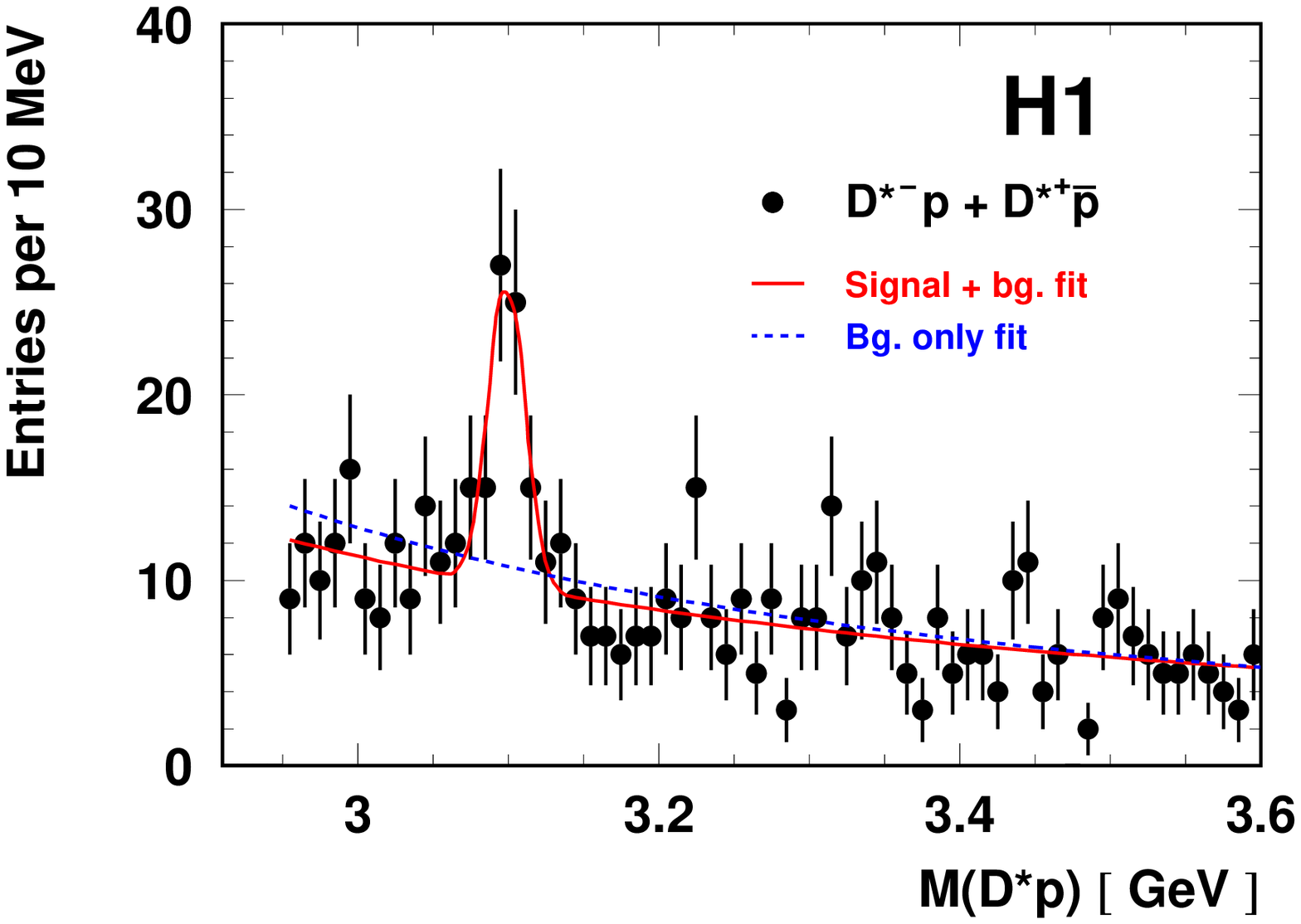,height=8cm}
  \epsfig{file=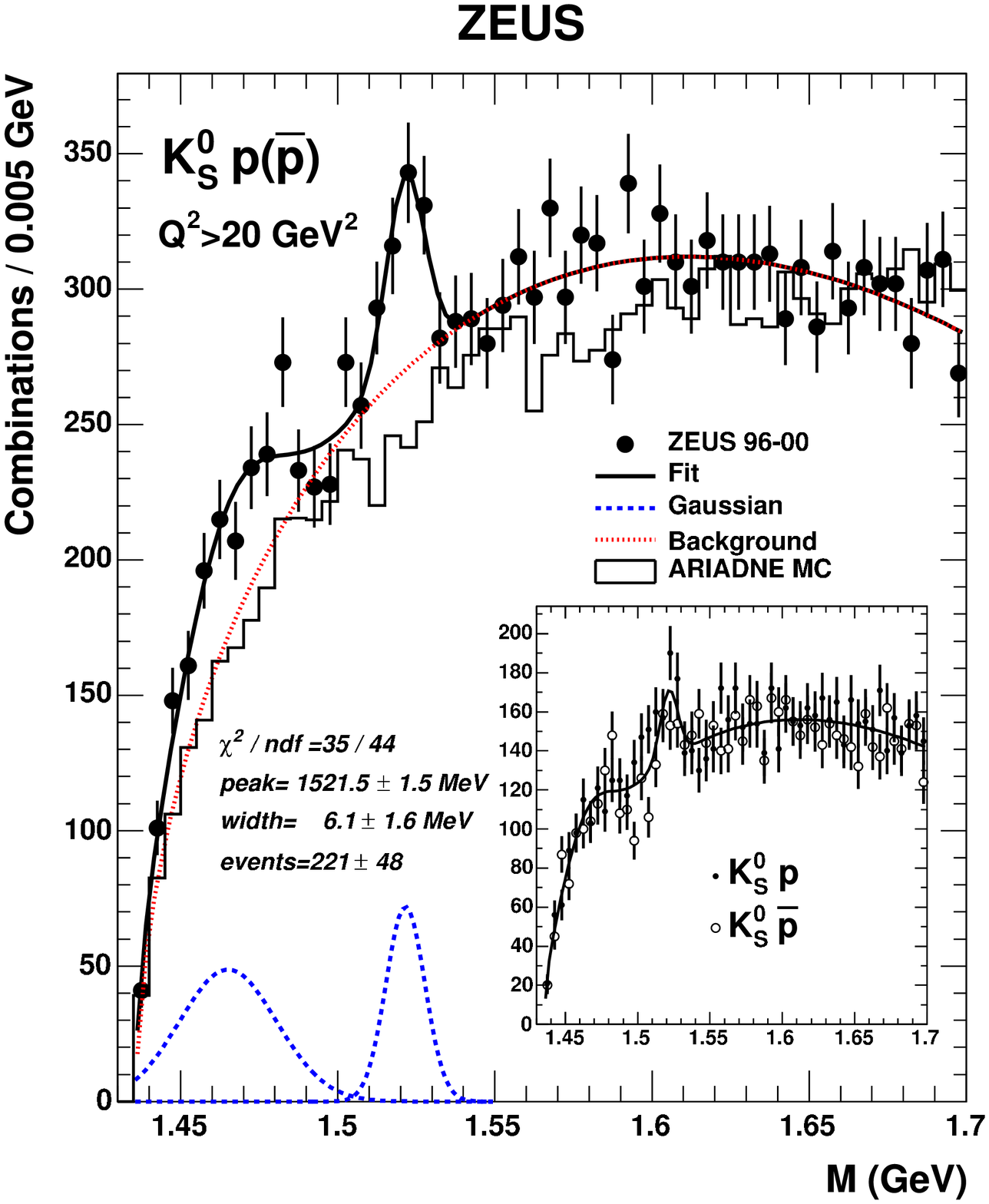,height=8cm}
  \caption{
    (Left) The H1 reconstructed mass distribution of $D^* p$ (Right)
    The ZEUS reconstructed mass distribution of $K^0_s p(\bar{p})$.}
  \label{fig:pq}
\end{figure}

Both H1 and ZEUS have recently reported evidence for
observations\cite{Aktas:2004qf,Karshon:2004kt} of pentaquark states.
ZEUS observe a peak (see figure \ref{fig:pq} (right)) in the
reconstructed mass distribution of $K^0_s p(\bar{p})$ which they
interpret as being the decay product of a $\Theta^+ (uudd\bar{s})$.
The kaon is identified from its decay to $\pi^+\pi^-$ and the proton
from $dE/dx$ measurements. The mass peak is found at 1521.5 $\pm$ 1.5
(stat.) +2.8/-1.7 (syst.) MeV with a Gaussian width of 6.1 $\pm$ 1.5
MeV compatible with the experimental resolution. Meanwhile H1 observe
a peak (see figure \ref{fig:pq} (left)) in the reconstructed mass
distribution of $D^* p$ which they interpret as being the decay
product of a $\Theta^0_c (uudd\bar{c})$. The $D^*$ is identified from
its ``golden decay channel'' $D^{*+} \rightarrow D^0\pi^+_s
\rightarrow K^-\pi^+\pi^+_s$ and the proton from $dE/dx$ measurements.
The mass peak is found at 3099 $\pm$ 3 (stat.) $\pm$ 5 (syst.)  MeV
with a Gaussian width of 12 $\pm$ 3 MeV compatible with the
experimental resolution. Both experiments calculate a probability of
background fluctuations causing these signals of around 5$\sigma$.
However each experiment is yet to confirm the other's observation.

\section{Conclusions}
\label{sec:conc}
HERA, the only electron-proton collider in the world, plays a unique
role in global particle physics. Its $ep$ collision experiments, H1
and ZEUS, can study the structure of the proton with unprecedented
precison, as revealed by measurements of the neutral and charged
current cross section measurements. Furthermore, H1 and ZEUS are able
to search for evidence of physics beyond the Standard Model in new
areas of phase space and indeed fascinating hints of potential signals
are seen.

\section[]{Acknowledgements}
Many thanks to Stephan Narison and the local organisiors and
participants for such an interesting and enjoyable conference.

\end{document}